\documentclass[prl,superscriptaddress,aps,floatfix,amsmath,amssymb]{revtex4}

\begin{document}

%\catchline{}{}{}{}{}
%
%%%%%%%%%%%%%%%%%%%%%%%%%%%%%%%%%%%%%%%%%%%%%%%%%%%%%%%%%%%%%%%%%%%%

\title{Classical Vs Quantum correlations in composite systems}

\author{Luigi Amico}
\address{CNR-MATIS-IMM $\&$ Dipartimento di Fisica e Astronomia Universit\'a di Catania\\
via S. Sofia 64
95123 Catania, Italy}
\affiliation{Center for Quantum Technologies, National University of Singapore,\\
	117542 Singapore, Singapore}

\author{Sougato Bose}

\address{Department of Physics and Astronomy, University College London\\
Gower St., London WC1E 6BT, UK}

\author{Vladimir E. Korepin}

\address{C. N. Yang Institute for Theoretical Physics, State University of New York at Stony Brook, \\ NY 11794-3840, USA}

\author{Vlatko  Vedral}

\address{Center for Quantum Technologies, National University of Singapore,\\
	117542 Singapore, Singapore}
\address{Department of Physics, National University of
	Singapore, \\ 2 Science Drive 3, Singapore 117542}
\address{Department of Physics, University of Oxford, Clarendon Laboratory, Oxford, \\ OX1 3PU, UK}

%\begin{history}
%\received{Day Month Year}
%\comby{(xxxxxxxxxx)}
%\end{history}

\begin{abstract}
Here we provide the contributions' abstracts  published in  a volume we edited as a special issue in  International Journal of Modern Physics B. 
The volume deals with the  recent progress in quantifying quantum correlations beyond the generic notion of 'correlations  in a quantum system'. The main goal of the special issue is to provide authoritative  reviews on
selected topics  discussed in the field in the last few years.
Nevertheless many articles contain significant original research.  The
opening section of the issue is dedicated to foundational aspects of
quantum mechanics.  The second section deals with diverse quantum
information insights into the analysis of quantum resources (like
entanglement and quantum discord). Many of the concepts presented in the first part of the issue are applied
to spin systems in the third section. The last section of the issue is focused on the dynamics of quantum discord in open systems and in quantum communication protocols.

\end{abstract}

%\keywords{Keyword1; keyword2; keyword3.}

%
\maketitle

\newpage

\vspace*{5cm}

\section{{\large Preface}}
\vspace*{2cm}

Quantum mechanics was discovered about hundred years ago. It provides the ex- planation of a variety of different phenomena in Nature. Nevertheless, a number of foundational aspects are still debated. In the last few years, quantum information has made a remarkable contribution to analyzing many foundational problems whose relevance, in turn, have been boosted by the progress in quantum technologies. Quantum information shed new light also on quantum statistical physics and condensed matter. It is of central importance to understand quantitatively the interplay between classical and quantum physics. The formal point of view stating that classical physics can be recovered from quantum mechanics when Plank constant goes to zero turns out not to be sufficient. Indeed, both quantum and classical correlations are present at fixed values of PlankÕs constant.
The present focus issue deals with the recent progress in quantifying quantum correlations beyond the generic notion of Òcorrelations in a quantum systemÓ. Entanglement is the most famous such genuine quantum correlation. In the last few years however, it has been realized that disentangled states may still contain genuine quantum correlations reflecting the superposition principle of quantum mechanics. This has led to the notion of quantum discord. Entanglement and quantum discord are resources that future quantum computers will exploit. Therefore it is important to have a picture of their dynamics, both in closed and open quantum systems. The analysis of entanglement and quantum discord provides useful insights in extended (many body) systems complementing the more traditional approaches based on energy spectrum analysis, transport properties, etc. In this context it is interesting to study how entanglement and discord propagate through the system, both for a basic physical understanding of the system and for applications.

The main goal of the issue is to provide authoritative reviews on selected topics discussed in the field in the last few years. Nevertheless many articles contain significant original research. The opening section of the issue is dedicated to foundational aspects of quantum mechanics. The second section deals with diverse quantum in- formation insights into the analysis of quantum resources (like entanglement and quantum discord). Many of the concepts presented in the first part of the issue are applied to spin systems in the third section. The last section of the issue is focused on the dynamics of quantum discord in open systems and in quantum communication protocols.
\\
\\
\\
{Luigi Amico}
\\
{Sougato Bose}
\\
{Vladimir E. Korepin} 
\\
{Vlatko Vedral}

%\tableofcontents

\newpage 

 \section{I: Foundational aspects of correlations in quantum mechanics}
{\small 
\begin{itemize}
\item
 {\bf  Classical, Quantum and Superquantum Correlations.} 
 \bigskip
 
GIANCARLO GHIRARDI \\
Department of Physics, University of Trieste, the Abdus Salam ICTP, Trieste, Strada Costiera 11, I-34151 Trieste, Italy 
\vspace{0.15cm}

\noindent
RAFFAELE ROMANO\\
Department of Physics, University of Trieste, the Abdus Salam ICTP, Trieste, Strada Costiera 11, I-34151 Trieste, Italy
\bigskip

{\it Abstract} \\
A deeper understanding of the origin of quantum correlations is expected to shed light on the physical principles underlying quantum mechanics. In this work, we reconsider the possibility of devising ``crypto-nonlocal theories", using a terminology firstly introduced by Leggett. We generalize and simplify the investigations on this subject which can be found in the literature. At their deeper level, such theories allow nonlocal correlations which can overcome the quantum limit.
\bigskip 

Read More: http://www.worldscientific.com/doi/abs/10.1142/S0217979213450112 \\
\hspace*{1.7cm}  http://arxiv.org/abs/1205.1162

\vspace*{1cm}

%\noindent
\item
{\bf Are quantum states real?}
 \bigskip 
 
LUCIEN HARDY \\
Perimeter Institute, 31 Caroline Street North, Waterloo, Ontario N2L 2Y5, Canada
\bigskip

{\it Abstract} \\
In this paper we consider theories in which reality is described by some underlying variables. Each value these variables can take represents an ontic state (a particular state of reality). The preparation of a quantum state corresponds to a distribution over the ontic states. If we make three basic assumptions, we can show that the distributions over ontic states corresponding to distinct pure states are non-overlapping. This means that we can deduce the quantum state from a knowledge of the ontic state. Hence, if these assumptions are correct, we can claim that the quantum state is a real thing (it is written into the underlying variables that describe reality). The key assumption we use in this proof is ontic indifference - that quantum transformations that do not affect a given pure quantum state can be implemented in such a way that they do not affect the ontic states in the support of that state. In fact this assumption is violated in the Spekkens toy model (which captures many aspects of quantum theory and in which different pure states of the model have overlapping distributions over ontic states). This paper proves that ontic indifference must be violated in any model reproducing quantum theory in which the quantum state is not a real thing. The argument presented in this paper is different from that given in a recent paper by Pusey, Barrett, and Rudolph. It uses a different key assumption and it pertains to a single copy of the system in question. 
\bigskip 

Read More: http://www.worldscientific.com/doi/abs/10.1142/S0217979213450124 \\
\hspace*{1.7cm} http://arxiv.org/abs/1205.1439

\vspace*{1cm}

%\noindent
 \item
 {\bf Observer Invariance of the Collapse Postulate of Quantum Mechanics}.
 \bigskip
 
MILTON A. DA SILVA, JR. \\
Centro de Cincias Naturais e Humanas, Universidade Federal do ABC, Rua Santa AdŽlia 166, 09210-170, Santo AndrŽ, S‹o Paulo, Brazil 
\vspace{0.15cm}

\noindent
ROBERTO M. SERRA \\
Centro de Cincias Naturais e Humanas, Universidade Federal do ABC, Rua Santa AdŽlia 166, 09210-170, Santo AndrŽ, S‹o Paulo, Brazil 
\vspace{0.15cm}

\noindent
LUCAS C. C\`ELERI \\
Centro de Cincias Naturais e Humanas, Universidade Federal do ABC, Rua Santa AdŽlia 166, 09210-170, Santo AndrŽ, S‹o Paulo, Brazil
\bigskip

{\it  Abstract} \\
We analyze the wave function \textit{collapse} as seem by two distinct observers (with identical detectors) in relative motion. Imposing that the measurement process demands information transfer from the system to the detectors, we note that although different observers will acquire different amount of information from their measurements due to correlations between spin and momentum variables, all of them will agree about the orthogonality of the outcomes, as defined by their own reference frame. So, in this sense, such a quantum mechanical postulate is observer invariant, however the \textit{effective efficiency of the measurement process} differs for each observer.
\bigskip
 
Read More: http://www.worldscientific.com/doi/abs/10.1142/S0217979213450136 \\
\hspace*{1.7cm}  http://arxiv.org/abs/1012.4447	

\vspace*{1cm}

%\noindent
 \item
 {\bf Coherent states and the classical-quantum limit considered from the point of view of entanglement}
\bigskip

THOMAS DURT\\
 Institut Fresnel, Domaine Universitaire de Saint-J\`erome, Avenue Escadrille Normandie-NiŽmen, 13397 Marseille Cedex 20, France 
 \vspace{0.15cm}

\noindent
VINCENT DEBIERRE \\
Institut Fresnel, Domaine Universitaire de Saint-J\`erome, Avenue Escadrille Normandie-NiŽmen, 13397 Marseille Cedex 20, France
\bigskip

{\it Abstract} \\
We consider the quasi-classical situation in which a quantum system interacts with another quantum system or with a quantum environment without getting entangled with it. We show how this regime is intimately linked to three paradigms commonly used in classical, pre-quantum physics to describe particles (that is, the material point, the test particle and the diluted particle (droplet model)).
This entanglement-free regime also provides a simplified insight on what is called in the decoherence approach ``islands of classicality'', that is, preferred bases that would be selected through evolution by a Darwinist mechanism which aims at optimizing information. We show how, under very general conditions, coherent states are natural candidates for classical pointer states. This occurs essentially because, when a (supposedly bosonic) system coherently exchanges only one quantum at a time with its (supposedly bosonic) environment, coherent states of the system do not get entangled with the environment, due to the bosonic symmetry.
\bigskip

Read More: http://www.worldscientific.com/doi/abs/10.1142/S0217979213450148 \\
\hspace*{1.7cm}

\vspace*{1cm}

%\noindent
\item
{\bf  Correlations of decay times of entangled composite unstable systems}.
\bigskip

 THOMAS DURT\\
Institut Fresnel, Domaine Universitaire de Saint-J\`erome, Avenue Escadrille Normandie-NiŽmen, 13397 Marseille Cedex 20, France 
\bigskip

{\it  Abstract} \\
The role played by Time in the quantum theory is still mysterious by many aspects. In particular it is not clear today whether the distribution of decay times of unstable particles could be described by a Time Operator. As we shall discuss, different approaches to this problem (one could say interpretations) can be found in the literature on the subject. As we shall show, it is possible to conceive crucial experiments aimed at distinguishing the different approaches, by measuring with accuracy the statistical distribution of decay times of entangled particles. Such experiments can be realized in principle with entangled kaon pairs.
\bigskip

Read More: http://www.worldscientific.com/doi/abs/10.1142/S021797921345015X \\
\hspace*{1.7cm} http://arxiv.org/abs/1206.5715

\vspace*{1cm}

%\noindent
\item
{\bf  Are quantum correlations genuinely quantum?}
\bigskip

ANTONIO DI LORENZO \\
Instituto de F'sica, Universidade Federal de Uberlandia, Av. Jo‹o Naves de Avila 2121, Uberlandia, Minas Gerais, 38400-902, Brazil
\bigskip

{\it Abstract} \\
It is shown that the probabilities for the spin singlet can be reproduced through classical resources, with no communication between the distant parties, by using merely shared (pseudo-)randomness. If the parties are conscious beings aware of both the hidden-variables and the random mechanism, then one has a conspiracy. If the parties are aware of only the random variables, they may be induced to believe that they are able to send instantaneous information to one another. It is also possible to reproduce the correlations at the price of reducing the detection efficiency. It is further demonstrated that the same probability decomposition could be realized through action-at-a-distance, provided it existed.
\bigskip

Read More: http://www.worldscientific.com/doi/abs/10.1142/S0217979213450161 \\
\hspace*{1.7cm} http://arxiv.org/abs/1205.0878

\end{itemize}

}

\section{II: Information theory approach to correlations}
{\small
\begin{itemize}
\item
{\bf  Correlations in Quantum Physics} 
\bigskip

ROSS DORNER \\
Blackett Laboratory, Imperial College London, Prince Consort Road, London SW7 2AZ, UK \\
Clarendon Laboratory, University of Oxford, Parks Road, Oxford OX1 3PU, UK 
\vspace{0.15cm}

\noindent
VLATKO VEDRAL\\
Clarendon Laboratory, University of Oxford, Parks Road, Oxford OX1 3PU, UK \\
Centre for Quantum Technologies, National University of Singapore, 3 Science Drive 2, 117543, Singapore \\
Department of Physics, National University of Singapore, 2 Science Drive 3, 117542, Singapore 
\bigskip

{\it Abstract}\\
We provide an historical perspective of how the notion of correlations has evolved within quantum physics. We begin by reviewing Shannon's information theory and its first application in quantum physics, due to Everett, in explaining the information conveyed during a quantum measurement. This naturally leads us to Lindblad's information theoretic analysis of quantum measurements and his emphasis of the difference between the classical and quantum mutual information. After briefly summarizing the quantification of entanglement using these and related ideas, we arrive at the concept of quantum discord that naturally captures the boundary between entanglement and classical correlations. Finally we discuss possible links between discord and the generation of correlations in thermodynamic transformations of coupled harmonic oscillators.
\bigskip

Read More: http://www.worldscientific.com/doi/abs/10.1142/S0217979213450173\\
\hspace*{1.7cm} http://arxiv.org/abs/1208.4961

\vspace*{1cm}

%\noindent
\item
{\bf Entanglement and its multipartite extensions} 
\bigskip

ANDREAS OSTERLOH \\
FakultŠt fŸr Physik, Universit\"at Duisburg-Essen, Campus Duisburg, Lotharstr. 1, D-47048 Duisburg, Germany

\bigskip

{\it Abstract}\\
The aspects of many particle systems as far as their entanglement is concerned is highlighted. To this end we briefly review the bipartite measures of entanglement and the entanglement of pairs both for systems of distinguishable and indistinguishable particles. The analysis of these quantities in
macroscopic systems shows that close to quantum phase transitions, the entanglement of many particles typically dominates that of pairs. This
leads to an analysis of a method to construct many-body entanglement measures. SL-invariant measures are a generalization to quantities as
the concurrence, and can be obtained with a formalism containing two (actually three) orthogonal anti-linear operators. The main drawback of this anti-anti-linearlinear framework, namely to measure these quantities in the experiment,
is resolved by a formula linking the anti-linear formalism to an
equivalent linear framework.
\bigskip 

Read More: http://www.worldscientific.com/doi/abs/10.1142/S0217979213450185 \\
\hspace*{1.7cm}

\vspace*{1cm}

%\noindent
\item
{\bf (Quantumness in the context of) Resource Theories} 
\bigskip 

MICHAL HORODECKI \\
Institute for Theoretical Physics and Astrophysics, University of Gda\`nsk, Gda\`nsk, Poland
\vspace{0.15cm}

\noindent
National Quantum Information Centre of Gda\`nsk, Sopot, Poland\\
JONATHAN OPPENHEIM\\
Department of Physics and Astronomy, University College London, London, United Kingdom
\bigskip 

{\it Abstract}\\
We review the basic idea behind resource theories, where we quantify quantum resources by specifying a restricted class of operations.
This divides the state space into various sets, including states which are free (because they can be created under the
class of operations), and those which are a resource (because they cannot be). One can quantify
the worth of the resource by the relative entropy distance to the set of free states, and under certain conditions, this is a unique
measure which quantifies the rate of state to state transitions.
The framework includes entanglement, asymmetry and purity theory. It also includes thermodynamics, which is a hybrid resource theory
combining purity theory and asymmetry. Another hybrid resource theory which merges purity theory and
entanglement can be used to study quantumness of correlations and discord, and we present quantumness in this more
general framework of resource theories.
\bigskip

Read More: http://www.worldscientific.com/doi/abs/10.1142/S0217979213450197 \\
\hspace*{1.7cm} http://arxiv.org/abs/1209.2162

\vspace*{1cm}

%\noindent
\item
{\bf Theoretical insights on measuring quantum correlations} 
\bigskip 

DAVIDE GIROLAMI \\
School of Mathematical Sciences, The University of Nottingham, University Park, Nottingham NG7 2RD, UK
 \vspace{0.15cm}

\noindent
RUGGERO VASILE \\
School of Mathematical Sciences, The University of Nottingham, University Park, Nottingham NG7 2RD, UK 
\vspace{0.15cm}

\noindent
GERARDO ADESSO \\
School of Mathematical Sciences, The University of Nottingham, University Park, Nottingham NG7 2RD, UK
\bigskip

{\it Abstract}\\
We review a recently developed theoretical approach to the experimental detection and quantification of bipartite quantum correlations between a qubit and a $d$ dimensional system. Specifically, introducing a properly designed measure $Q$, the presented scheme allows us to quantify general quantum correlations for arbitrary states of $2\otimes d$ systems without the need to fully reconstruct them by tomographic techniques. We take in exam the specifics of the required experimental architecture in nuclear magnetic resonance and optical settings. Finally we extend this approach to models of open system dynamics and discuss possible advantages and limitations in such a context.

\bigskip

Read More: http://www.worldscientific.com/doi/abs/10.1142/S0217979213450203 \\
\hspace*{1.7cm} http://arxiv.org/abs/1208.5964

\vspace*{1cm}

%\noindent
\item
{\bf  The balance of quantum correlations for a class of feasible tripartite continuous variable states}
\bigskip 

STEFANO OLIVARES \\
Dipartimento di Fisica, Universitˆ degli Studi di Milano, I-20133 Milano, Italy \\
CNISM UdR Milano Statale, I-20133 Milano, Italy 
\vspace{0.15cm}

\noindent
MATTEO G. A. PARIS \\
Dipartimento di Fisica, Universitˆ degli Studi di Milano, I-20133 Milano, Italy \\
CNISM UdR Milano Statale, I-20133 Milano, Italy
\bigskip 

{\it Abstract} \\
We address the balance of quantum correlations for continuous variable (CV) states. In particular, we consider a class of feasible tripartite CV pure states and explicitly prove two Koashi-Winter-like conservation laws involving Gaussian entanglement of formation, Gaussian quantum discord and sub-system Von Neumann entropies. We also address the class of tripartite CV mixed states resulting from the propagation in a noisy environment, and discuss how the previous equalities evolve into inequalities.
\bigskip

Read More: http://www.worldscientific.com/doi/abs/10.1142/S0217979213450240 \\
\hspace*{1.7cm} http://arxiv.org/abs/1207.4370

\vspace*{1cm}

%\noindent
\item
{\bf Hierarchy of correlations via L\"uders measurements}
\bigskip 

SHUNLONG LUO \\
Academy of Mathematics and Systems Science, Chinese Academy of Sciences, Beijing 100190, People's Republic of China 
\vspace{0.15cm}

\noindent
SHUANGSHUANG FU \\
Academy of Mathematics and Systems Science, Chinese Academy of Sciences, Beijing 100190, People's Republic of China 
\bigskip 

{\it Abstract} \\
The classification and quantification of correlations (classical and quantum) in composite
quantum systems are of fundamental significance for quantum information processing.
While the paradigm of separability versus entanglement has been intensively
studied, the scenario of classicality versus quantumness, with focus on the quantum discord,
has also attracted many recent interests. In this paper, pursuing further the latter
scenario and exploiting the intrinsic structure of bipartite quantum states via local projective
measurements, we introduce the notion of coherent dimension of correlations in
terms of the L\"uders measurements. The coherent dimension can alternatively be regarded
as a generalization of the Schmidt number of a pure state. Furthermore, we propose some
families of measures for correlations, which extend naturally both the quantum discord
and the quantum mutual information (total correlations), and furthermore interpolate
between them. These quantities reveal some hierarchical structures, and provide a more
complete description, of both classical and quantum correlations in the quantum realm.
\bigskip 

Read More: http://www.worldscientific.com/doi/abs/10.1142/S0217979213450264 \\
\hspace*{1.7cm} 

\vspace*{1cm}

%\noindent
\item
{\bf Coherent and incoherent contents of correlations.} 
\bigskip 

KAVAN MODI\\
Clarendon Laboratory, Department of Physics, University of Oxford, Oxford, UK \\
Centre for Quantum Technologies, National University of Singapore, Singapore\\
MILE GU\\
Centre for Quantum Technologies, National University of Singapore, Singapore \\
\bigskip 

{\it Abstract} \\
We examine bipartite and multipartite correlations within the construct of unitary orbits. We show that the set of product states is a very small subset of set of all possible states, while all unitary orbits contain \emph{classically correlated} states. Using this we give meaning to degeneration of quantum correlations due to a unitary interactions, which we call coherent correlations. The remaining classical correlations are called incoherent correlations and quantified in terms of the distance of the joint probability distributions to its marginals. Finally, we look at how entanglement looks in this picture for the two-qubit case.
\bigskip

Read More: http://www.worldscientific.com/doi/abs/10.1142/S0217979213450276 \\
\hspace*{1.7cm} http://arxiv.org/abs/0902.0735

\vspace*{1cm}

%\noindent
\item
{\bf Matrix Product State representation for Slater Determinants and Configuration Interaction States} 
\bigskip 

PIETRO SILVI \\
Institut fŸr Quanteninformationsverarbeitung, UniversitŠt Ulm, D-89069 Ulm, Germany \\
International School for Advanced Studies (SISSA), Via Bonomea 265, I-34136 Trieste, Italy 
\vspace{0.15cm}

\noindent
DAVIDE ROSSINI\\
NEST, Scuola Normale Superiore and Istituto di Nanoscienze Ð CNR, Pisa, Italy 
\vspace{0.15cm}

\noindent
ROSARIO FAZIO \\
NEST, Scuola Normale Superiore and Istituto di Nanoscienze Ð CNR, Pisa, Italy \\
Center for Quantum Technologies, National University of Singapore, Republic of Singapore 
\vspace{0.15cm}

\noindent
GIUSEPPE E. SANTORO \\
International School for Advanced Studies (SISSA), Via Bonomea 265, I-34136 Trieste, Italy \\
International Centre for Theoretical Physics (ICTP), P. O. Box 586, I-34014 Trieste, Italy \\
CNR-IOM Democritos National Simulation Center, Via Bonomea 265, I-34136 Trieste, Italy 
\vspace{0.15cm}

\noindent
VITTORIO GIOVANNETTI \\
NEST, Scuola Normale Superiore and Istituto di Nanoscienze Ð CNR, Pisa, Italy \\
\bigskip 

{\it Abstract}\\
Slater determinants are product states of filled quantum fermionic orbitals. When they are expressed in a configuration space basis chosen a priori, their entanglement is bound and controlled. This suggests that an exact representation of Slater determinants as finitely-correlated states is possible. In this paper we analyze this issue and provide an exact Matrix Product representation for Slater determinant states. We also argue possible meaningful extensions that embed more complex configuration interaction states into the description.
\bigskip 

Read More: http://www.worldscientific.com/doi/abs/10.1142/S021797921345029X \\
\hspace*{1.7cm}  http://arxiv.org/abs/1205.4154

\end{itemize}

}

\section{III: Quantum versus classical correlations  in spin systems}
{\small
\begin{itemize}
\item
{\bf Quantum discord in the ground state of spin chains}
\bigskip 

MARCELO S. SARANDY\\
Instituto de F'sica, Universidade Federal Fluminense, Av. Gal. Milton Tavares de Souza s/n, Gragoat\`a, 24210-346, Niter\`oi, Rio de Janeiro, Brazil 
\vspace{0.15cm}

\noindent
THIAGO R. DE OLIVEIRA \\
Instituto de F'sica, Universidade Federal Fluminense, Av. Gal. Milton Tavares de Souza s/n, Gragoat\`a, 24210-346, Niter\`oi, Rio de Janeiro, Brazil 
\vspace{0.15cm}

\noindent
LUIGI AMICO \\
CNR-MATIS-IMM $\&$ Dipartimento di Fisica e Astronomia Universitˆ di Catania, via S. Sofia 64, 95123 Catania, Italy
\bigskip 

{\it Abstract}\\
The ground state of a quantum spin chain is a natural playground for investigating correlations. Nevertheless, not all correlations are genuinely of quantum nature. Here we
review the recent progress to quantify the quantumness of the correlations throughout
the phase diagram of quantum spin systems. Focusing to one spatial dimension, we dis-
cuss the behavior of quantum discord close to quantum phase transitions. In contrast to
the two-spin entanglement, pairwise discord is effectively long-ranged in critical regimes.
Besides the features of quantum phase transitions, quantum discord is especially feasible
to explore the factorization phenomenon, giving rise to nontrivial ground classical states
in quantum systems. The effects of spontaneous symmetry breaking are also discussed
as well as the identification of quantum critical points through correlation witnesses.
\bigskip 

Read More: http://www.worldscientific.com/doi/abs/10.1142/S0217979213450306 \\
\hspace*{1.7cm}  http://arxiv.org/abs/1208.4817

\vspace*{1cm}

%\noindent
\item
{\bf Interplay between quantum phase transitions and the behavior of quantum correlations at finite temperatures}
\bigskip 

T. WERLANG \\
Departamento de F'sica, Universidade Federal de S\~ao Carlos, S\~ao Carlos, SP 13565-905, Brazil 
\vspace{0.15cm}

\noindent
G. A. P. RIBEIRO \\
Departamento de F'sica, Universidade Federal de S\~ao Carlos, S\~ao Carlos, SP 13565-905, Brazil 
\vspace{0.15cm}

\noindent
GUSTAVO RIGOLIN \\
Departamento de F'sica, Universidade Federal de S\~ao Carlos, S\~ao Carlos, SP 13565-905, Brazil
\bigskip 

{\it Abstract} \\
We review the main results and ideas showing that quantum correlations at finite temperatures (T), in particular quantum discord, are useful tools in characterizing quantum phase transitions that only occur, in principle, at the unattainable absolute zero temperature. We first review some interesting results about the behavior of thermal quantum discord for small spin-1/2 chains and show that they already give us important hints of the infinite chain behavior. We then study in detail and in the thermodynamic limit (infinite chains) the thermal quantum correlations for
the XXZ and XY models, where one can clearly appreciate
that the behavior of thermal quantum discord at finite T is a useful tool to spotlight the critical point of a quantum phase transition.
\bigskip

Read More: http://www.worldscientific.com/doi/abs/10.1142/S021797921345032X \\
\hspace*{1.7cm}  http://arxiv.org/abs/1205.1046

\vspace*{1cm}

%\noindent
\item
{\bf Quantum discord and related measures of quantum correlations in finite XY chains}
\bigskip

N. CANOSA \\
Departamento de F'sica-IFLP-CONICET-CIC, Universidad Nacional de La Plata, C.C.67, La Plata (1900), Argentina 
\vspace{0.15cm}

\noindent
L. CILIBERTI \\
Departamento de F'sica-IFLP-CONICET-CIC, Universidad Nacional de La Plata, C.C.67, La Plata (1900), Argentina 
\vspace{0.15cm}

\noindent
R. ROSSIGNOLI \\
Departamento de F'sica-IFLP-CONICET-CIC, Universidad Nacional de La Plata, C.C.67, La Plata (1900), Argentina
\bigskip

{\it Abstract}\\
We examine the quantum correlations of spin pairs in the ground state of finite $XY$ chains in a transverse field, by evaluating the quantum discord as well as other related entropic measures of quantum correlations. A brief review of the latter, based on generalized entropic forms, is also included. It is shown that parity effects are of crucial importance for describing the behavior of these measures below the critical field. It is also shown that these measures reach full range in the immediate vicinity of the factorizing field, where they become independent of separation and coupling range. Analytical and numerical
results for the quantum discord, the geometric discord and other measures in spin chains with nearest neighbor coupling and in fully connected spin arrays are also provided.
\bigskip

Read More: http://www.worldscientific.com/doi/abs/10.1142/S0217979213450331 \\
\hspace*{1.7cm} http://arxiv.org/abs/1206.2995

\vspace*{1cm}

%\noindent
\item
{\bf Genuine correlations in finite-size spin systems}
\bigskip 

GIAN LUCA GIORGI \\
Department of Physics, University College Cork, Cork, Republic of Ireland
\vspace{0.15cm}

\noindent
THOMAS BUSCH \\
Department of Physics, University College Cork, Cork, Republic of Ireland \\
Quantum Systems Unit, Okinawa Institute of Science and Technology, Okinawa 904-0411, Japan

\bigskip 

{\it Abstract}\\
Genuine multipartite correlations in finite-size XY chains are studied as a function of the applied external magnetic field. We find that, for low temperatures, multipartite correlations are sensitive to the parity change in the Hamiltonian ground state, given that they exhibit a minimum every time that the ground state becomes degenerate. This implies that they can be used to detect the factorizing point, that is, the value of the external field such that, in the thermodynamical limit, the ground state becomes the tensor product of single-spin states.
\bigskip 

Read More: http://www.worldscientific.com/doi/abs/10.1142/S0217979213450343 \\
\hspace*{1.7cm} http://arxiv.org/abs/1206.1726

\vspace*{1cm}

%\noindent
\item
{\bf Transport of Quantum Correlations across a spin chain}
\bigskip 

TONY J. G. APOLLARO\\
Dipartimento di Fisica, Universitˆ della Calabria, 87036 Arcavacata di Rende (CS), Italy \\
INFN Ð Gruppo collegato di Cosenza, Italy 
\vspace{0.15cm}

\noindent
SALVATORE LORENZO\\
Dipartimento di Fisica, Universitˆ della Calabria, 87036 Arcavacata di Rende (CS), Italy \\
INFN Ð Gruppo collegato di Cosenza, Italy 
\vspace{0.15cm}

\noindent
FRANCESCO PLASTINA \\
Dipartimento di Fisica, Universitˆ della Calabria, 87036 Arcavacata di Rende (CS), Italy \\
INFN Ð Gruppo collegato di Cosenza, Italy

\bigskip 

{\it Abstract}\\
Some of the recent developments concerning the propagation of quantum correlations across spin channels are reviewed. In particular, we focus on the improvement of the transport
efficiency obtained by the manipulation of few energy parameters (either end-bond strengths or local magnetic fields) near the sending and receiving sites. We give a physically insightful description of various such schemes and discuss the transfer of both entanglement and of quantum discord.
\bigskip

Read More: http://www.worldscientific.com/doi/abs/10.1142/S0217979213450355 \\
\hspace*{1.7cm} http://arxiv.org/abs/1207.6048

\vspace*{1cm}

%\noindent
\item
{\bf Quenching Dynamics and Quantum Information}
\bigskip 

TANAY NAG \\
Department of Physics, Indian Institute of Technology Kanpur, Kanpur 208016, India 
\vspace{0.15cm}

\noindent
AMIT DUTTA\\
Department of Physics, Indian Institute of Technology Kanpur, Kanpur 208016, India 
\vspace{0.15cm}

\noindent
AYOTI PATRA \\
Department of Physics, University of Maryland, College Park, Maryland 20742-4111, USA

\bigskip 

{ \it Abstract}\\
We review recent studies on the measures of zero temperature quantum correlations namely, the quantum entanglement (concurrence) and discord present
in the final state of a transverse $XY$ spin chain following a quench through quantum critical points;
the aim of these studies is to explore the scaling of the above quantities as a function of the quench rate.
A comparative study between the concurrence and the quantum discord shows that their behavior is qualitatively
the same though there are quantitative differences. For the present model, the scaling of both the
quantities are given by the scaling of the density of the defect present in the final state though one
can not find a closed form expression for the discord. We also extend our study of quantum discord to a transverse Ising chain in the presence of
a three spin interaction. Finally, we present a study of the dynamical evolution of quantum discord and concurrence when two central qubits, initially prepared in a Werner state, are coupled to the environmental $XY$ spin chain which is driven through quantum critical points. The qualitative behavior of quantum discord and concurrence are found to be similar as that of the decoherence factor.
\bigskip 

Read More: http://www.worldscientific.com/doi/abs/10.1142/S0217979213450367 \\
\hspace*{1.7cm} http://arxiv.org/abs/1206.0559

\vspace*{1cm}

%\noindent
\item
{\bf Fluctuations assisted stationary entanglement in driven quantum systems}
\bigskip 

DIMITRIS ANGELAKIS \\
Science Department, Technical University of Crete, 73100 Chania, Greece 
\vspace{0.15cm}

\noindent
STEFANO MANCINI \\
School of Science and Technology, University of Camerino, 62032 Camerino, Italy 
\bigskip 

{ \it Abstract}\\
We analyze quantum correlations arising in two coupled dimer systems in the presence of
independent losses and driven by a fluctuating field. For the case of the interaction being of a Heisenberg exchange type, we  first analytically show the possibility for stationary entanglement and then analyze its robustness as a function of the signal-to-noise ratio of the pump. We  find that for a common  fluctuating driving  field, stochastic resonance effects appears as function of the ratio between  field strength and noise strength. The
effect disappears in the case of uncorrelated or separate pumps. Our result is general
and could be applied to different quantum systems ranging from electron spins in solid
state, to ions trap technologies and cold atom set ups.
\bigskip 

Read More: http://www.worldscientific.com/doi/abs/10.1142/S0217979213450379 \\
\hspace*{1.7cm} http://arxiv.org/abs/1107.0905

\end{itemize}

}

\section{IV: Quantum Communication and dynamics of quantum discord in open systems}

{\small

\begin{itemize}
\item

{\bf Quantum discord as a resource in quantum communication}
\bigskip 

VAIBHAV MADHOK \\
Department of Physics and Computer Science, Wilfrid Laurier University, Waterloo, Ontario N2L 3C5, Canada \\
Center for Quantum Information and Control, University of New Mexico, Albuquerque, NM 87131-0001, USA 
\vspace{0.15cm}

\noindent
ANIMESH DATTA \\
Clarendon Laboratory, Department of Physics, University of Oxford, OX1 3PU, United Kingdom

\bigskip 

{\it Abstract} \\
As quantum technologies move from the issues of principle to those of practice, it is important
to understand the limitations on attaining tangible quantum advantages. In the realm of quantum communication,
quantum discord captures the damaging effects of a decoherent environment. This is a
consequence of quantum discord quantifying the advantage of quantum coherence in quantum communication.
This establishes quantum discord as a resource for quantum communication processes.
We review this progress, which derives a quantitative relation between the yield of the fully quantum
Slepian-Wolf protocol in the presence of noise and the quantum discord of the state involved. The
significance of quantum discord in noisy versions of teleportation, super-dense coding, entanglement
distillation and quantum state merging are discussed. These results lead to open questions regarding
the tradeoff between quantum entanglement and discord in choosing the optimal quantum states for
attaining palpable quantum advantages in noisy quantum protocols.
\bigskip 

Read More: http://www.worldscientific.com/doi/abs/10.1142/S0217979213450410 \\
\hspace*{1.7cm} http://arxiv.org/abs/1204.6042

\vspace*{1cm}

%\noindent
\item
{\bf Quantum discord, decoherence and quantum phase transition}
\bigskip 

INDRANI BOSE \\
Department of Physics, Bose Institute 93/1, Acharya Prafulla Chandra Road, Kolkata Ð 700009, India 
\vspace{0.15cm}

\noindent
AMIT KUMAR PAL \\
Department of Physics, Bose Institute 93/1, Acharya Prafulla Chandra Road, Kolkata Ð 700009, India 
\bigskip 

{ \it Abstract}\\
Quantum discord is a more general measure of quantum correlations than entanglement 
and has been proposed as a resource in certain quantum information processing
tasks. The computation of discord is mostly confined to two-qubit systems for which an
analytical calculation scheme is available. The utilization of quantum correlations in
quantum information-based applications is limited by the problem of decoherence, i.e.,
the loss of coherence due to the inevitable interaction of a quantum system with its
environment. The dynamics of quantum correlations due to decoherence may be studied
in the Kraus operator formalism for different types of quantum channels representing
system-environment interactions. In this review, we describe the salient features of the
dynamics of classical and quantum correlations in a two-qubit system under Markovian
(memoryless) time evolution. The two-qubit state considered is described by the reduced
density matrix obtained from the ground state of a spin model. The models considered
include the transverse-field XY model in one dimension, a special case of which is the
transverse-field Ising model, and the XXZ spin chain. The quantum channels studied include the amplitude damping, bit-flip, bit-phase-flip and phase-flip channels. The Kraus
operator formalism is briefly introduced and the origins of different types of dynamics discussed. One can identify appropriate quantities associated with the dynamics of
quantum correlations which provide signatures of quantum phase transitions in the spin
models. Experimental observations of the different types of dynamics are also mentioned.
\bigskip 

Read More: http://www.worldscientific.com/doi/abs/10.1142/S0217979213450422 \\
\hspace*{1.7cm} http://arxiv.org/abs/1205.1300

\vspace*{1cm}

\item
{\bf Dynamics of quantum correlations in two-qubit systems within non-markovian environments}
\bigskip 

ROSARIO LO FRANCO \\
CNISM and Dipartimento di Fisica, Universitˆ di Palermo, via Archirafi 36, 90123 Palermo, Italy 
\vspace{0.15cm}

\noindent
BRUNO BELLOMO \\
UniversitŽ Montpellier 2, Laboratoire Charles Coulomb UMR 5221, F-34095, Montpellier, France 
\vspace{0.15cm}

\noindent
SABRINA MANISCALCO \\
Turku Center for Quantum Physics, Department of Physics and Astronomy, University of Turku, FIN20014, Turku, Finland \\
SUPA, EPS Physics, Heriot-Watt University, Edinburgh, EH14 4AS, UK
\vspace{0.15cm}

\noindent
GIUSEPPE COMPAGNO \\
CNISM and Dipartimento di Fisica, Universitˆ di Palermo, via Archirafi 36, 90123 Palermo, Italy
\bigskip 

{ \it Abstract}\\
Knowledge of the dynamical behavior of correlations with no classical counterpart, like entanglement, nonlocal correlations and quantum discord, in open quantum systems is of primary interest because of the possibility to exploit these correlations for quantum information tasks. Here we review some of the most recent results on the dynamics of correlations in bipartite systems embedded in non-Markovian environments that, with their memory effects, influence in a relevant way the system dynamics and appear to be more fundamental than the Markovian ones for practical purposes. Firstly, we review the phenomenon of entanglement revivals in a two-qubit system for both independent environments and a common environment. We then consider the dynamics of quantum discord in non-Markovian dephasing channel and briefly discuss the occurrence of revivals of quantum correlations in classical environments.
\bigskip 

Read More: http://www.worldscientific.com/doi/abs/10.1142/S0217979213450537 \\
\hspace*{1.7cm} http://arxiv.org/abs/1205.6419

\vspace*{1cm}

%\noindent

\item
{\bf Quantum discord under system-environment coupling: the two-qubit case}
\bigskip 

JIN-SHI XU \\
Key Laboratory of Quantum Information, University of Science and Technology of China, CAS, Hefei 230026, People's Republic of China 
\vspace{0.15cm}

\noindent
CHUAN-FENG LI \\
Key Laboratory of Quantum Information, University of Science and Technology of China, CAS, Hefei 230026, People's Republic of China
\bigskip 

{ \it Abstract}\\
Open quantum systems have attracted great attention, since inevitable coupling between
quantum systems and their environment greatly  affects the features of interest of these
systems. Quantum discord, is a measure of the total nonclassical correlation in a quantum system that includes, but is not exclusive to, the distinct property of quantum
entanglement. Quantum discord can exist in separated quantum states and plays an important role in many fundamental physics problems and practical quantum information
tasks. There have been numerous investigations on quantum discord and its counterpart
classical correlation. This short review focuses on highlighting the system-environment
dynamics of two-qubit quantum discord and the influence of initial system-environment correlations on the dynamics of open quantum systems. The external control effect on
the dynamics of open quantum systems are involved. Several related experimental works
are discussed.
\bigskip

 Read More: http://www.worldscientific.com/doi/abs/10.1142/S0217979213450549 \\
\hspace*{1.7cm} http://arxiv.org/abs/1205.0871

\vspace*{1cm}

%\noindent

\item
{\bf  System-Spin Environment Dynamics of Quantum Discord}
\bigskip 

BEN-QIONG LIU \\ 
Key Laboratory of Neutron Physics and Institute of Nuclear Physics and Chemistry, China Academy of Engineering Physics, Mianyang 621900, China 
\vspace{0.15cm}

\noindent
BIN SHAO \\
Key Laboratory of Cluster Science of Ministry of Education and School of Physics, Beijing Institute of Technology, Beijing 100081, China 
\vspace{0.15cm}

\noindent
JUN-GANG LI \\
Key Laboratory of Cluster Science of Ministry of Education and School of Physics, Beijing Institute of Technology, Beijing 100081, China 
\vspace{0.15cm}

\noindent
JIAN ZOU \\
Key Laboratory of Cluster Science of Ministry of Education and School of Physics, Beijing Institute of Technology, Beijing 100081, China 
\bigskip 

{ \it Abstract}\\
As a fundamental concept in quantum mechanics, entanglement has been widely studied in various fields of quantum physics mainly because of its possible applications in
quantum information. While interest remains strong, recent work has relatively comprehensively studied nonclassical correlations beyond entanglement which provide computational speedup and quantum enhancement even in separable states. Different indicators
of nonclassical correlations have been proposed; among them the quantum discord has
received special attention, which is supposed to account for all the nonclassical correlations present in a bipartite state, including entanglement. Almost all quantum states
contain non vanishing quantum discord, and it has been increasing in relevance in quantum computing, quantum communication, quantum phase transitions, etc. For realistic
quantum systems, their inevitable coupling with the complex environment leads to fast
decoherence rates. Decoherence means the destruction of quantum coherence, known as
the process by which quantum information is degraded. It plays a crucial role in understanding the quantum to classical transition. Interestingly, under some conditions,
quantum discord is completely unaffected by the decoherence environment during a long
time interval, suggesting that quantum discord is robust to external perturbations and
may introduce a new breakthrough to quantum technologies such as quantum computers.
\bigskip 

Read More: http://www.worldscientific.com/doi/abs/10.1142/S0217979213450550 \\
\hspace*{1.7cm}

\end{itemize}
}
\end{document}